\documentclass[conference]{IEEEtran} % 10 pt default?
\IEEEoverridecommandlockouts
\usepackage{cite}
\usepackage{amsmath,amssymb,amsfonts}
\usepackage{algorithmic}
\usepackage{graphicx}
\usepackage{textcomp}
\usepackage{xcolor}

\usepackage{geometry}
\def\BibTeX{{\rm B\kern-.05em{\sc i\kern-.025em b}\kern-.08em
    T\kern-.1667em\lower.7ex\hbox{E}\kern-.125emX}}

\begin{document}

\title{Process Visualization of Manufacturing Execution System (MES) Data\\

\thanks{This work was carried out within the Confirm Smart Manufacturing Centre (https://confirm.ie/) funded by Science Foundation Ireland (grant number: 16/RC/3918).}
}

\author{\IEEEauthorblockN{Meadhbh O'Neill}
\IEEEauthorblockA{\textit{Confirm} \\
\textit{University of Limerick}\\
Limerick, Ireland \\
meadhbh.oneill@ul.ie}
\and
\IEEEauthorblockN{Jeff Morgan}
\IEEEauthorblockA{\textit{Confirm} \\
\textit{National University of Ireland}\\
Galway, Ireland \\
jeff.morgan@nuigalway.ie}
\and
\IEEEauthorblockN{Kevin Burke}
\IEEEauthorblockA{\textit{Confirm} \\
\textit{University of Limerick}\\
Limerick, Ireland \\
kevin.burke@ul.ie}
}

\maketitle

\begin{abstract}
Process visualizations of data from manufacturing execution systems (MESs) provide the ability to generate valuable insights for improved decision-making. Industry 4.0 is awakening a digital transformation where advanced analytics and visualizations are critical. Exploiting MESs with data-driven strategies can have a major impact on business outcomes. The advantages of employing process visualizations are demonstrated through an application to real-world data. Visualizations, such as dashboards, enable the user to examine the performance of a production line at a high level. Furthermore, the addition of interactivity facilitates the user to customize the data they want to observe. Evidence of process variability between shifts and days of the week can be investigated with the goal of optimizing production.
\end{abstract}

\begin{IEEEkeywords}
process visualization, manufacturing execution system, real-world data, business intelligence, dashboard
\end{IEEEkeywords}

%%=====================================================%%
%%=====================================================%%
\section{Introduction}\label{sec:introduction}
A manufacturing execution system (MES) has the functionality to support manufacturing execution processes from the stage of production order release until the delivery of the finished goods. This computerized system documents and monitors all of the components involved in the transformation of raw materials to the final product. The MES provides a data management system, which can be used to generate comprehensive reports of all features associated with the process \cite{industry 4.0 background}. In the current digital transformation era of Industry 4.0, it is necessary for businesses to consider their place in this paradigm shift, where high levels of digital capabilities have an essential role in the factories of the future. Manufacturing execution systems (MESs) can aid businesses in achieving their Industry 4.0 goals through increased operation visibility and traceability \cite{mes_ind_4_1}. However, the mainstream use of MES data for improving process performance is somewhat unexplored, and such data has not historically been used in academia \cite{mes_ind_4_2}.

With the continuous expansion of data availability, advanced analytics may be used to extract value from the data. Analytics can exploit the generation of large volumes of data from these powerful industrial systems by investigating the performance of the manufacturing processes, quality of products and supply chain optimization. Examining historical data can identify inefficiencies and enable corrective or preventative steps to be carried out. Data-driven strategies are critical to support the business in optimizing its performance by gathering and analyzing data throughout the processes. Businesses can manage uncertainties in their operations through process mining \cite{toolbox, process mining mapping, workload}, artificial intelligence (AI) and machine learning (ML) \cite{decision_making}. Businesses desire systems that can not only provide efficient overall production performance, but also systems that are reactive to real time information streams \cite{coupling}. MESs can be integrated with predictive production planning and predictive maintenance, where anomalies can be detected and faults can be predicted \cite{ml, predictive_analysis}. The latest MESs are capable of autonomously executing schedules and reacting to changes in production due to unexpected disturbances \cite{cyber-physical}. In order to increase the decision-making capabilities of such systems, collaborative frameworks between centralized scheduling systems and MESs have been introduced \cite{collab}. Autonomic MESs are capable of addressing highly dynamic and uncertain situations through adaptability, autonomy and flexibility \cite{autonomic}.

Data analytics is the foundation of these recent advances. When handling real-world data, exploratory analysis is an essential first step before implementing any AI or ML tools. The volume, variety and velocity of data that are generated by large scale, complex systems demands an in-depth examination to ensure that the processes are understood fully before any advanced analytics are carried out. Business intelligence (BI) tools provide effective methods for visualizing data in a user-friendly way. Indeed, the use of data visualization --- especially in the form of interactive dashboards --- can provide the end user with a method for better understanding the process being measured. Businesses can view and interact with their data in different ways and visualize large batches of data in order to generate valuable insights. These tools can be utilized as decision support systems, where the user can quickly and effectively make informed business decisions assisted by data.

The key elements of dashboards are described in Section~\ref{sec:dashboards}. In Section~\ref{sec:data_challenges}, we highlight various data challenges that a business may encounter while analyzing their data. The advantages of utilizing process visualizations are presented in Section~\ref{sec:application_to_real_world_data}, where real-world MES data is examined. Finally, we close with some discussion in Section~\ref{sec:discussion}.

%%=====================================================%%
%%=====================================================%%
\section{Dashboards}\label{sec:dashboards}
\begin{figure}[b!]
\centerline{\includegraphics[width = 0.8\columnwidth]{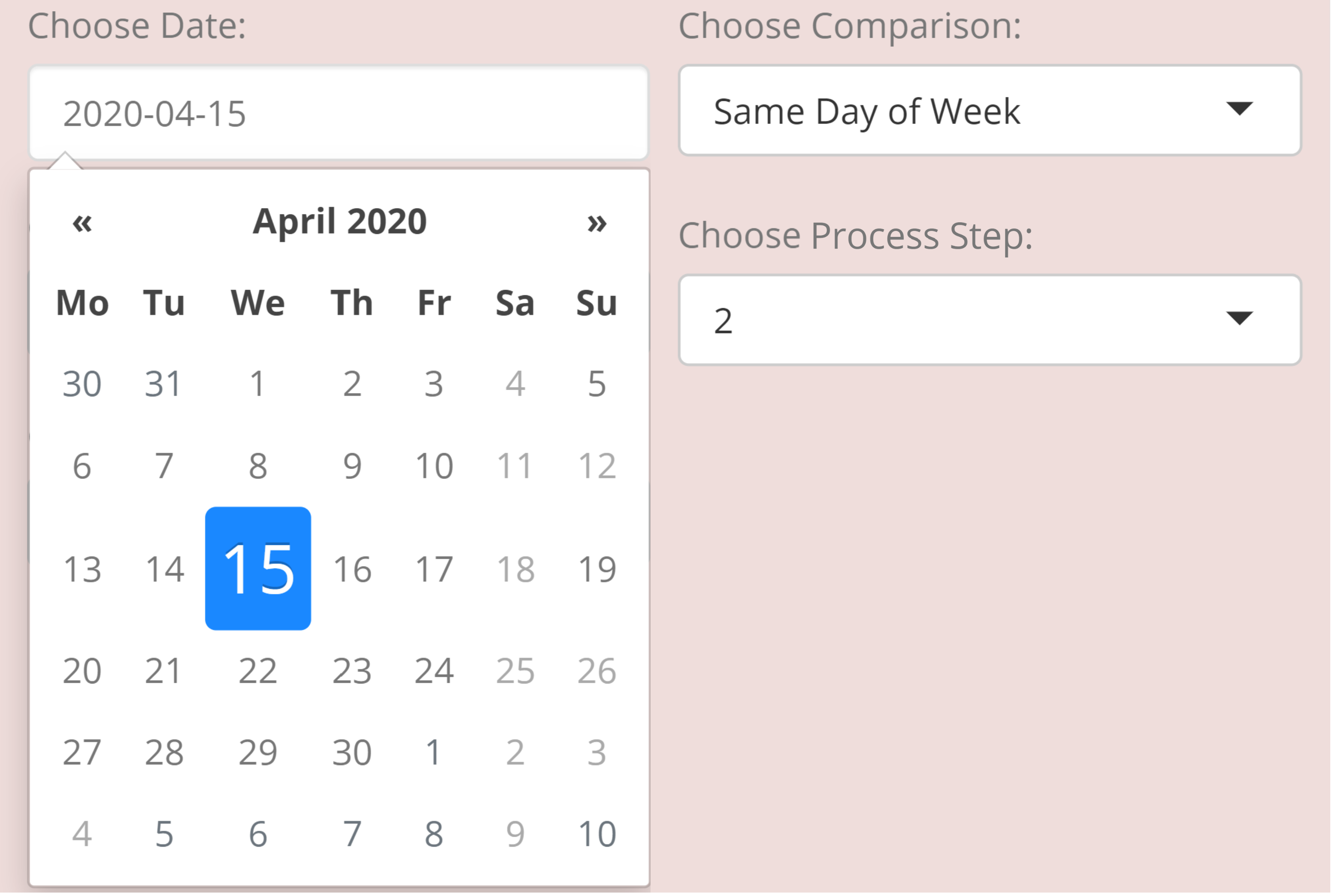}}
\caption{User defined inputs in homepage.}
\label{fig:figs_inputs}
\end{figure}

Dashboards, such as Tableau \cite{tableau} and PowerBI \cite{powerBI}, are useful BI tools that provide a clear perspective of the data and help to improve business processes. R Shiny \cite{R_Core, R_Studio, shiny, mastering_shiny} is a popular and our preferred method of building BI tools, as it packages smart process analytics into a web application. This combines both analytics and decision-making to generate actionable business insights. The interface allows the user to change input values and manipulate parameters through text boxes, filters and sliders. This facilitates the user to customize the data they want to see. Fig.~\ref{fig:figs_inputs} provides an example of this, where the user can choose specific steps within the process to view using a drop-down list. Specific days can also be selected using the interactive interface. An additional level of interactivity can be achieved by creating plots with plotly \cite{plotly}, whereby the user can hover over different points to display coordinate values and other information. The user can zoom in and closely examine different areas of the plot, while also having the ability to filter on particular groups using the legend.

%%=====================================================%%
% Fig: Inputs
%%=====================================================%%

\begin{figure}[t!]
% \vspace{0.5cm}
\centerline{\includegraphics[width = 0.6\columnwidth]{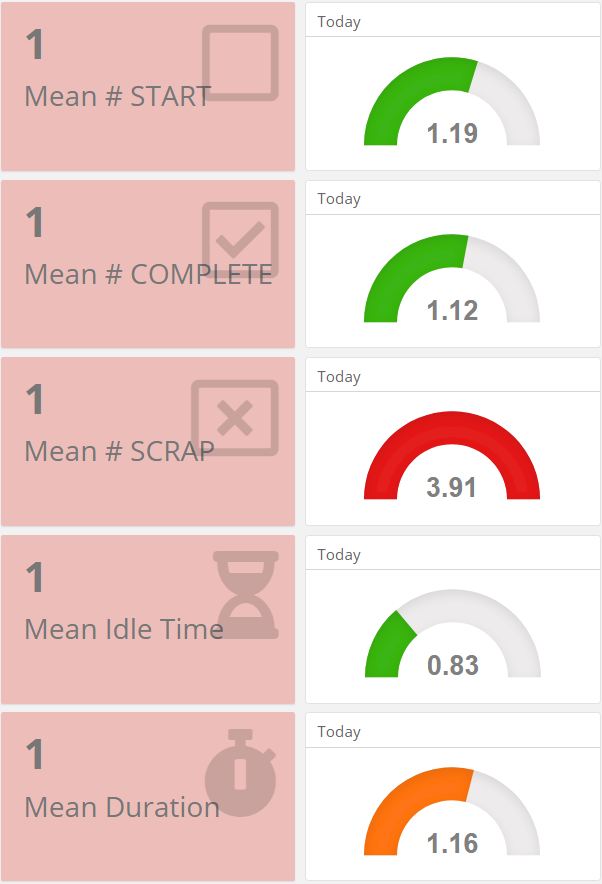}}
\caption{Example of the homepage where the performance of a process step on a chosen day can be viewed at a high level. The rules for coloring the gauges are outlined in Table~\ref{tab:color_rules}.}
\label{fig:figs_daily_performance}
\end{figure}

In addition to examining specific steps within a process, an overall homepage is also useful where high level information about all process steps can quickly be consumed by the user as shown in Fig.~\ref{fig:figs_daily_performance}. Displaying leading indicators where the performance of the production line can be observed at a high level is beneficial to understand how well it is operating. This provides the user with increased visibility into the operations and information so they can steer the process more effectively. Visualizations that show how the performance of the production line varies over time are informative. The average behavior and the variation of the processes can be utilized to identify unusual cases. Extreme observations can be investigated systematically by way of a root cause analysis. If it is the case that certain observations lie outside of the expected limits, it is possible to examine what is causing the process to be out-of-control. The dashboard can be automated to provide real time visibility, where it responds to changes in the data automatically. This allows the user to monitor the process in real time.

%%=====================================================%%
% Fig: Daily Performance
%%=====================================================%%

%%=====================================================%%
%%=====================================================%%
\section{Data Challenges}\label{sec:data_challenges}
Handling the data and understanding the specific characteristics of the data set is typically a substantial challenge. Working closely with subject matter experts is vital to contextualize the data and to gain an understanding of the processes that are taking place. It is generally not feasible to produce insights in a vacuum or without understanding the context of the data.

The volume of data that a business has to manage is ever increasing. For example, these data may have been collected for quality or reporting reasons. As these data are available, businesses begin to think about analyzing the data in order to identify key metrics and generate meaningful and actionable insights, although this is not the purpose for which the data were originally gathered; indeed, in some cases, data may have been collected over time but with no particular goal in mind. Analyzing the data is often an afterthought and so the true value of the data to the business is unknown. Ideally, a research question would be posed in advance of the data collection process, so that an appropriate study can be designed to ensure that the relevant data are collected. Adopting a structured approach such as the ``goal, question, metric" (GQM) paradigm \cite{gqm orig} could be implemented to improve manufacturing processes \cite{gqm}. For most businesses, this is not the case and the data that they have available may or may not possess the capability of generating desirable insights.

Challenges relating to data quality and integrity are also common. For example, some unit identification numbers may not be scanned or entered correctly, or the location of certain equipment could be mislabeled. Depending on how and where the data are stored, access and extraction may not be straightforward. Ensuring that the data are extracted in a homogenized form is essential. The business may have multiple databases, which store data in different formats. The content of these databases may change over time as, for example, new variables are now measured and others are no longer measured. There may also be formatting differences between data that are current/live and older data that have been moved to an archive. For data stored across multiple files, there may be small variations such as the name of a particular variable or its position in the data set. Data processing and cleaning is a key step to confirm that the raw data contain what is expected and are in a form amenable to analysis.

%%=====================================================%%
%%=====================================================%%
\section{Application to Real-World Data}\label{sec:application_to_real_world_data}
%%=====================================================%%
\subsection{Data}
\subsubsection{Description}
\begin{figure}[b!]
\centerline{\includegraphics[width = 0.8\columnwidth]{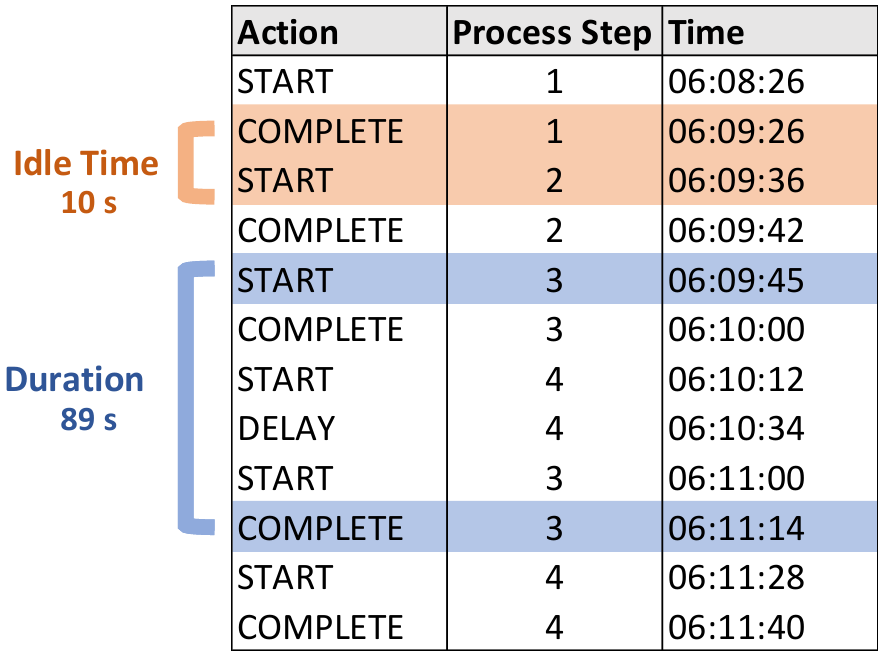}}
\caption{Raw process log data, including an example of the idle time and duration calculations.}
\label{fig:figs_idle_time_calc}
\end{figure}

The MES data examined here track the production of units through a series of seven process steps. The process logs within the MES contain a record of all actions and activities performed on the production line. An example of the raw process log data is available in Fig.~\ref{fig:figs_idle_time_calc}. There are several variables of interest, including the numbers of starts and completes in a particular time period, as well as the idle time between actions and the duration of a process step. The MES data analyzed runs over several months in 2020 and there are approximately 1.4 million observations.

%%=====================================================%%
% Figs: Idle time calc
%%=====================================================%%

\subsubsection{Variables}
There are four possible actions: ``start", ``complete", ``scrap" and ``delay". The start action indicates that a unit has entered a particular process step. Likewise, the complete action provides a record of the unit finishing at a specific process step. Scraps occur if there are issues with the units and they must be discarded. Delays signify problems with the units but not to the extent that they needed to be scrapped.

Idle time is the time between process steps during which no value is being added towards a final product. This is calculated at a unit level and is computed as the time from when the unit is finished at a step until it starts at the next step (see Fig.~\ref{fig:figs_idle_time_calc}).

The duration of time a unit spends at a process step is calculated as the time from the first start to the last complete. This approach captures any delays experienced by the unit, e.g., in Fig.~\ref{fig:figs_idle_time_calc}, the unit needed to revisit Process Step 4 before finally completing Process Step 3.

\subsubsection{Transformation}
In order to anonymize the results, all analysis has been completed using rescaled data. The tables and figures present data that have been divided by a mean value to create dimensionless values. Results can be interpreted as multiples relative to the mean value.

%%=====================================================%%
\subsection{Methods}\label{subsec:methods}
\subsubsection{Metrics}\label{subsubsec:metrics}
Metrics including the mean and standard deviation of the variables are used to capture the average behavior of the different process steps on the production line, along with an indication of how the performance varies over time. Additionally, percentiles are used throughout the analysis as a measure of the variation in the data. Percentiles are better suited in this case as they are more robust towards outliers and data that are skewed compared to bands produced using the standard deviation (typically based on normality assumptions). Percentiles give the value below which a proportion of the data falls. For example, the mean curve of the number of scraps over time can be computed with percentile bounds, which indicate the range in which 95\% of the values lie. The 95\% bounds are calculated by obtaining the range between the ${2.5}^\text{th}$ and the ${97.5}^\text{th}$ percentiles.

%%=====================================================%%
\subsubsection{Moving Average}\label{subsubsec:moving_average}
Using a moving average (MA) window is useful to smooth over noisy data in order to identify trends over time. For example, analyzing the number of starts for a given process step is an indicator of workload. If the raw number of starts in each 30-minute window is used, we have found the resulting visualizations (not shown) to be quite noisy. Therefore, we make use of a moving average of order three whereby we take the average of three 30-minute periods: the one of interest and ones on either side of this. We have found this to be sufficient for our purposes where local trends were of interest but some noise reduction was desirable. In contrast, a higher order moving average creates smoother curves but that would be more suited if global trends were of interest.

This MA technique of a rolling 30-minute interval on either side of the time window of interest is used to compute MA mean curves. Smooth percentile bounds indicating the range in which 95\% of the values from the three-time windows lie are also produced.

%%=====================================================%%
\subsubsection{Statistical Control}\label{subsubsec:statistical_control}
The MA mean curves and 95\% bounds for a process step can be computed and used as a template for an average day from a statistical control perspective. A selected day, such as ``today", can be taken and compared to the average behavior in order to assess the performance of a particular process step. For example, today's duration for this process step at different times of the day can be overlaid on the average curves. This provides an opportunity to evaluate how the process step is performing relative to the average, and also if the data lies within the 95\% bounds. If today's data are more extreme than the bounds, then this is a cause for concern that will require further investigation. Times of the day or shifts which vary greatly can be determined.

%%=====================================================%%
\subsection{Process Step Analysis}\label{subsec:workstation_level}
\subsubsection{Collective Data}\label{subsubsec:collective_data}
\begin{figure}[b!]
% \vspace{0.2cm}
\centerline{\includegraphics[width = \columnwidth]{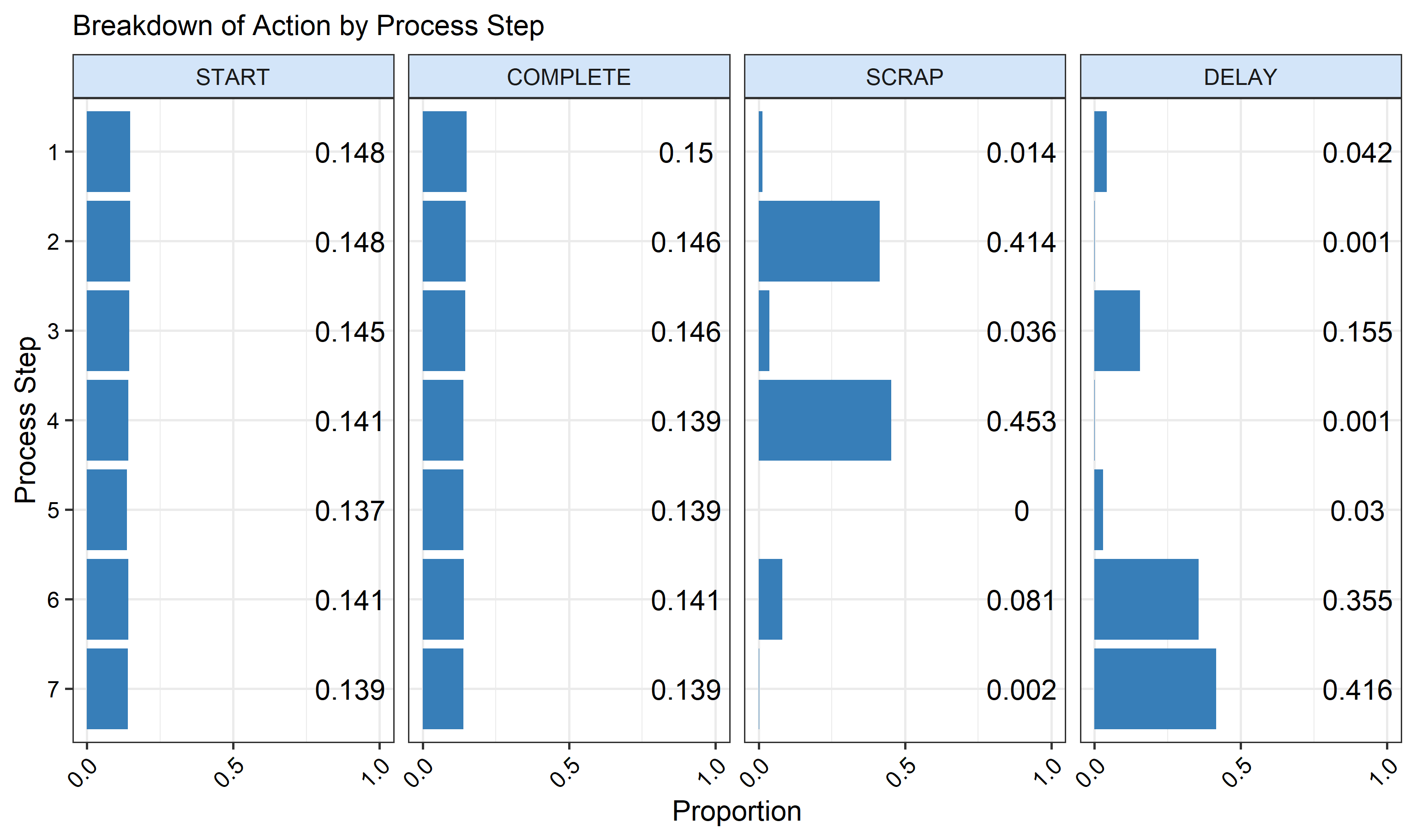}}
\caption{Breakdown of the proportion of action by process step.}
\label{fig:figs_breakdown}
\end{figure}

A breakdown of the main actions by Process Step is available in Fig.~\ref{fig:figs_breakdown}. The start and complete actions are relatively evenly split between the process steps. The process steps further along the production line tend to have slightly fewer starts. This is expected because, as a unit passes through more steps of the process, there are more chances for an issue to arise, e.g., machine failure or operator error. The majority of scraps occur at Process Step 2 (41\%) and Process Step 4 (45\%). Delays are most common at Process Step 6 (36\%) and Process Step 7 (42\%).

%%=====================================================%%
\subsubsection{Workload}\label{subsubsec:workload}
\begin{figure}[t!]
\centerline{\includegraphics[width = \columnwidth]{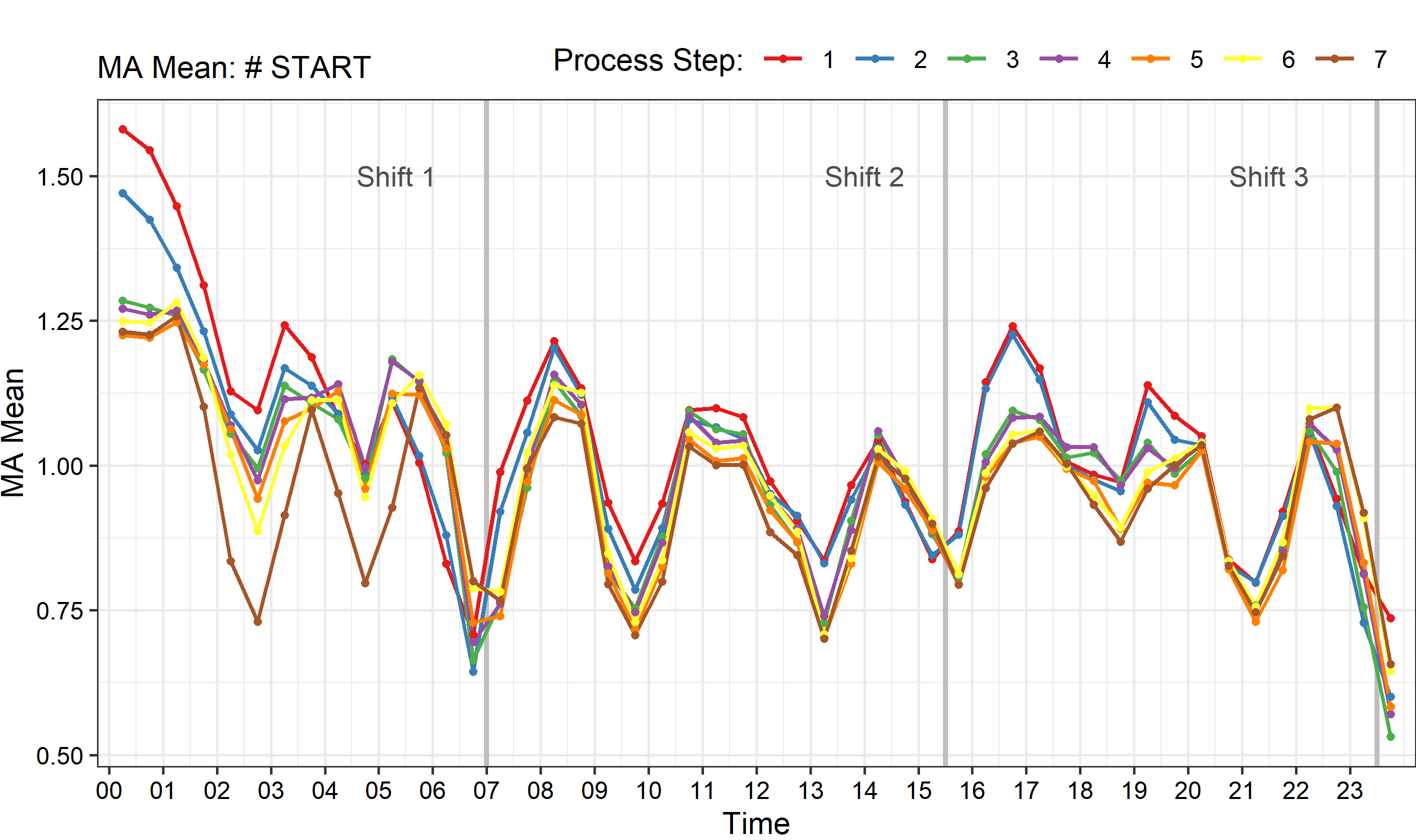}}
\caption{Number of starts (moving average) over time split by process steps.}
\label{fig:figs_ma_start}
\end{figure}

The MA mean number of starts over time (30-minute intervals) taken over all dates and split by process step is shown in Fig.~\ref{fig:figs_ma_start}. The time of day is separated by vertical gray lines indicating the shift type. All of the process steps behave similarly throughout the shifts. The volume of work tends to be best within the first hour of the shift commencing. This is perhaps due to an initial few hours of uninterrupted work before the operators start to take breaks. Interestingly, the highest number of starts for all process steps occurs during the early stage of Shift 1, which corresponds to midnight. For the remainder of the time, the MA mean number of starts varies between 0.75 and 1.25, where the evident cyclic nature aligns with common work/break periods.

\begin{figure}[t!]
\centerline{\includegraphics[width = \columnwidth]{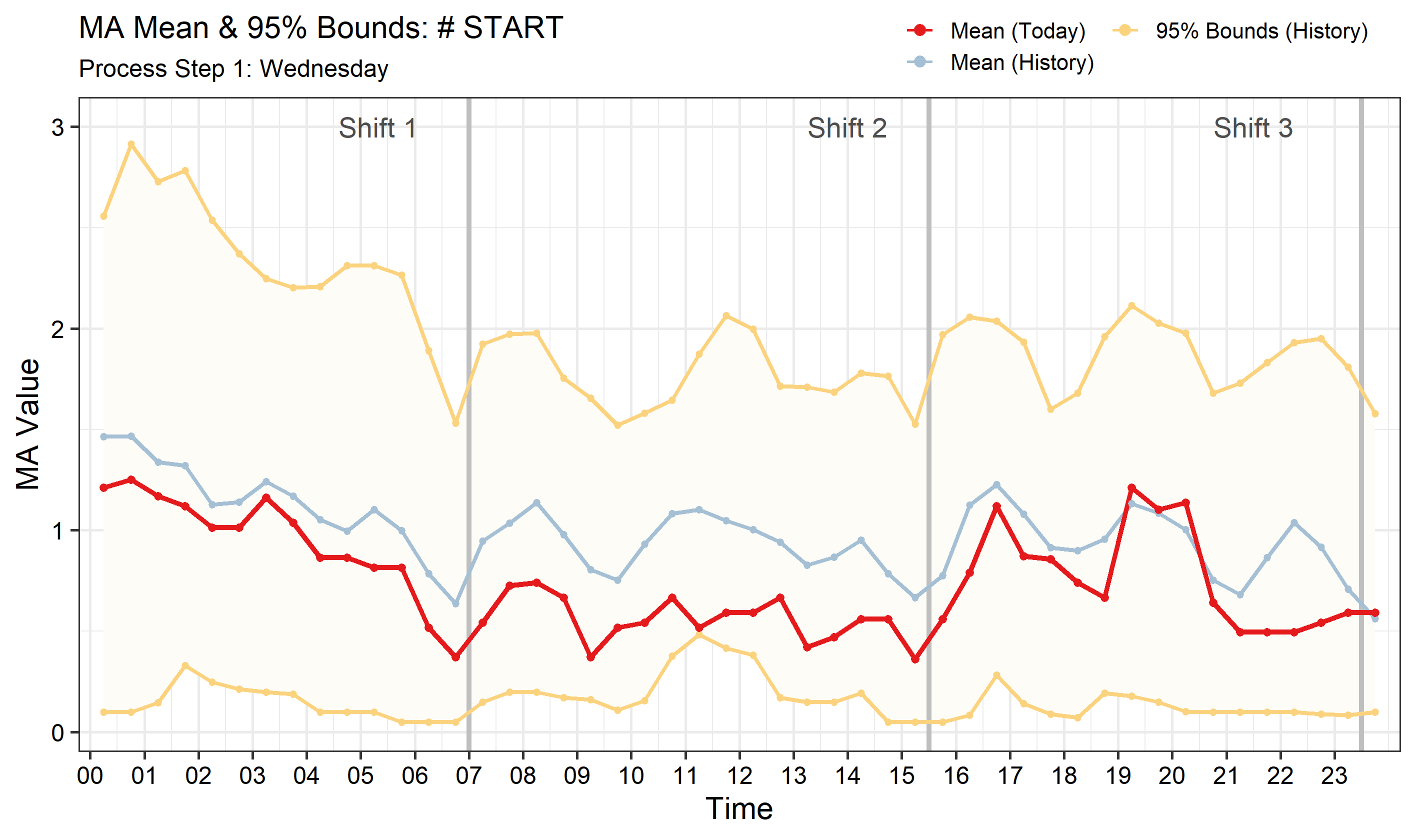}}
\caption{Example of using the MA mean and 95\% bounds to assess the performance of Process Step 1 ``today".}
\label{fig:figs_ma_start_today}
\end{figure}

It is more informative to incorporate uncertainty by displaying the 95\% bounds. Fig.~\ref{fig:figs_ma_start_today} focuses on Process Step 1 and plots the MA mean (blue line) and 95\% bounds (yellow lines) of the number of starts over time. This example takes the average over all Wednesdays. Examining specific days of the week is useful due to there being a difference in the hours worked on some days. These bounds show that there is a relatively high level of variability around the overall average (blue line), especially in Shift 1. Indeed, the mean line and bounds can be used as a template within which a ``usual" day's work should lie. Here ``today" is taken as a specific Wednesday in the data (red line). It generally tracks the historic mean line, albeit there are fewer starts at most time points with the largest discrepancies being in Shift 2. However, today's performance is not deemed ``unusual" as it lies within the 95\% bounds.

%%=====================================================%%
% Fig: breakdown
%%=====================================================%%

%%=====================================================%%
% Fig: ma_start
%%=====================================================%%

%%=====================================================%%
% Fig: ma_start_today
%%=====================================================%%

%%=====================================================%%
\subsection{Unit Analysis}\label{subsec:unit_level}
\subsubsection{Idle Time}\label{subsubsec:idle_time}
The average idle time where the unit is not in production is displayed in Table~\ref{tab:idle_time}. This includes idle time analysis metrics for each process step, including the mean, standard deviation (SD), median, ${2.5}^\text{th}$ percentile (P2.5) and the ${97.5}^\text{th}$ percentile (P97.5). The mean and median values differ greatly, which suggests that the idle time distributions are skewed with large outlying values. All of the process steps have a median idle time of less than 0.3. However, the idle time before Process Step 3 and Process Step 6 vary greatly with upper bounds of 15.15 and 6.44 respectively.

\begin{figure}[b!]
\centerline{\includegraphics[width = \columnwidth]{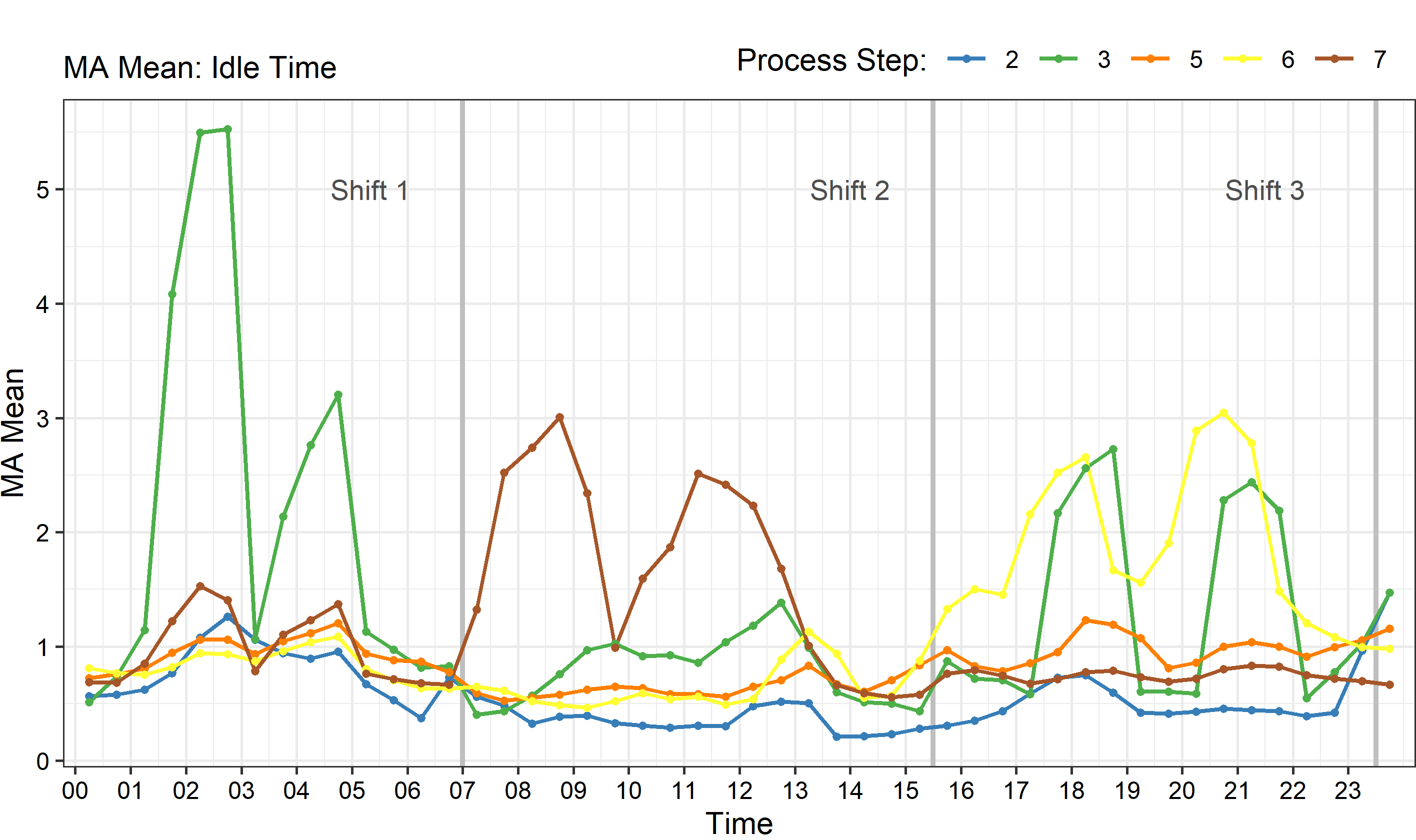}}
\caption{Moving average mean of idle time split by process step over time.}
\label{fig:figs_idle_time}
\end{figure}

The MA mean idle time over time taken over all dates and split by process step is shown in Fig.~\ref{fig:figs_idle_time}. Process Step 3 has large mean idle time values relative to the other steps during Shift 1. Process Step 7 has large values during Shift 2, while Process Step 6 has larger values in Shift 3. The mean idle time for the remaining process steps behave similarly to each other throughout time.

%%=====================================================%%
% Table: Idle Time
%%=====================================================%%
\begin{table}[b!]
\caption{Idle Time Analysis}
\begin{center}
\begin{tabular}{|c|cc|ccc|}
\hline
\textbf{Process Step}&\textbf{Mean}&\textbf{SD}&\textbf{Median}&\textbf{P2.5}&\textbf{P97.5} \\
\hline
2 & 0.55 & 6.20 & 0.02 & 0.02 & 4.34 \\ 
  3 & 1.35 & 10.29 & 0.02 & 0.02 & 15.15 \\ 
  5 & 0.84 & 3.69 & 0.26 & 0.03 & 4.56 \\ 
  6 & 1.12 & 5.99 & 0.20 & 0.02 & 6.44 \\ 
  7 & 1.16 & 5.67 & 0.30 & 0.12 & 4.39 \\ 
\hline
\end{tabular}
\label{tab:idle_time}
\end{center}
\end{table}

%%=====================================================%%
% Fig: Idle time
%%=====================================================%%

%%=====================================================%%
\subsubsection{Duration}\label{subsubsec:duration}
The average time that a unit spends at each process step is presented in Table~\ref{tab:duration}. There are some differences between the mean and median values, particularly for Process Step 1 and Process Step 2. The distributions of duration values for the other process steps are less skewed, with mean and median values that are relatively close. Process Steps 2 and 4 take the longest on average to complete. As identified in Fig.~\ref{fig:figs_breakdown}, these process steps account for over 85\% of the scraps observed in the data. The large upper bound value for Process Step 1 is due to delays that are infrequent (see Fig.~\ref{fig:figs_breakdown}) but large when they occur. The upper bounds for the remaining process steps are approximately less than 0.25 in duration.

\begin{figure}[t!]
\centerline{\includegraphics[width = \columnwidth]{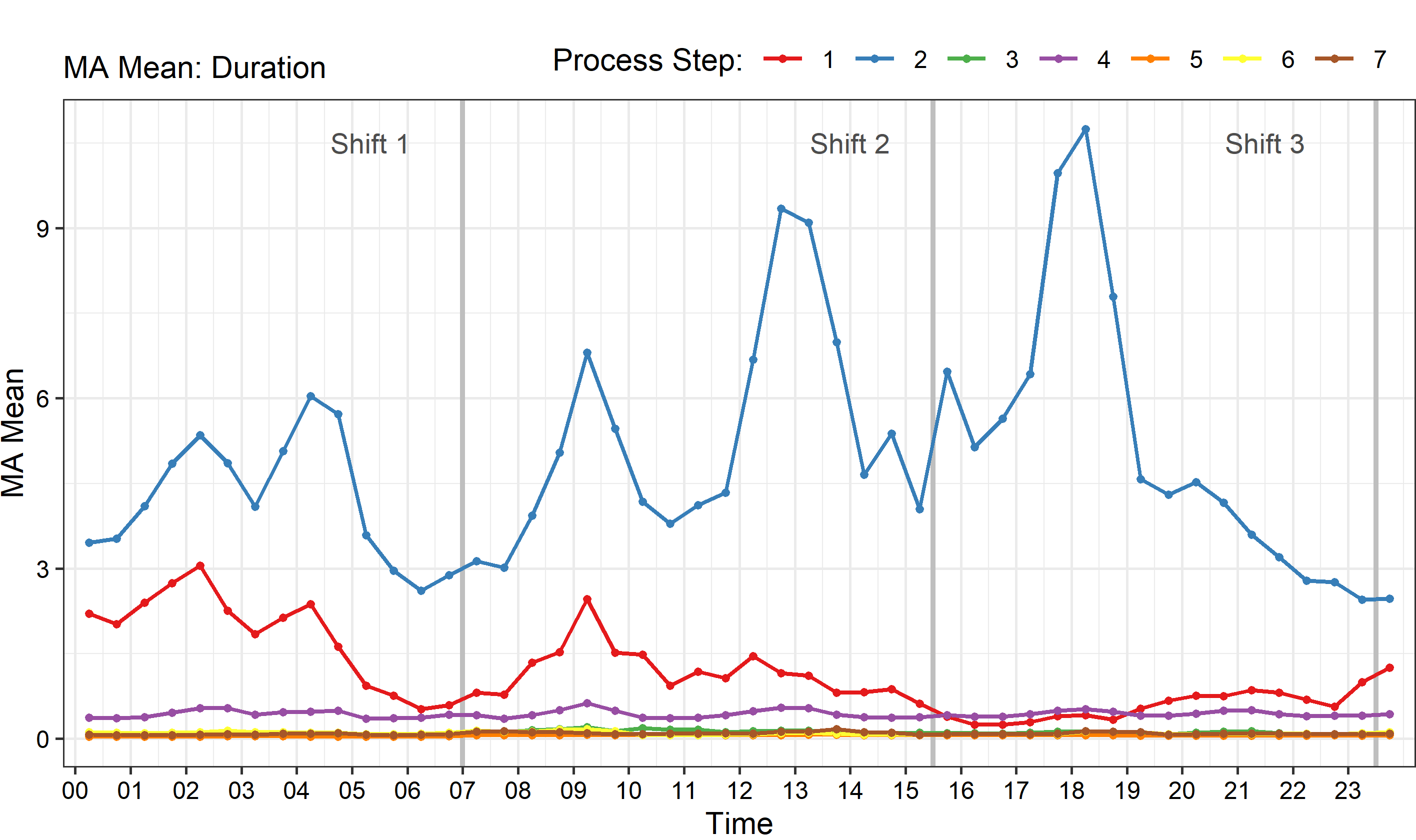}}
\caption{Moving average mean of duration split by process step over time.}
\label{fig:figs_duration}
\end{figure}

The MA mean duration over time taken over all dates and split by process step is shown in Fig.~\ref{fig:figs_duration}. Process Step 2 has large mean duration values relative to the other steps throughout time. This value appears to increase from Shift 1 to Shift 3. There are some large duration values for Process Step 1 during Shift 1. This seems to settle in Shift 2 and 3. There are no very large mean duration values for the other process steps.

%%=====================================================%%
\subsection{Overall Daily Dashboard}\label{subsec:daily_performance}
Static graphics such as those previously displayed are useful, but some additional interactivity can be included, for example, by allowing the user to select particular days and/or process steps themselves.

The dynamic dashboard features a homepage where the performance of each process step can be observed at a high level. The user can customize the display by choosing the date, the type of comparison they wish to make (overall data or same day of the week), and the process step as illustrated in Fig.~\ref{fig:figs_inputs}. This homepage comprises of a set of gauges that change color depending on how well the process step is operating relative to the historical data. A snapshot of the homepage is captured in Fig.~\ref{fig:figs_daily_performance}. In this case, ``today" is a specific Wednesday in the data. Process Step 2 is being examined and the comparison is made relative to all other Wednesdays in the data. The performance metrics computed are the number of starts, completions and scraps, and the mean idle time and duration. The boxes containing gauges display values for today. The color of the gauge depends on where today's value lies relative to the distribution of values from the historical data. The rules for coloring the gauges are outlined in Table~\ref{tab:color_rules}. In this example, there are green gauges for the number of starts and completions and the average idle time for the day. This indicates that these elements of the process step are performing well. The length of time a unit spends at the process step is longer relative to the historical data as highlighted by the orange duration gauge. On the particular day shown in Fig.~\ref{fig:figs_daily_performance}, there were a large number of scraps resulting in a red gauge.

%%=====================================================%%
% Table: Duration
%%=====================================================%%
\begin{table}[t!]
\caption{Duration Analysis}
\begin{center}
\begin{tabular}{|c|cc|ccc|}
\hline
\textbf{Process Step}&\textbf{Mean}&\textbf{SD}&\textbf{Median}&\textbf{P2.5}&\textbf{P97.5} \\
\hline
1 & 1.21 & 5.18 & 0.10 & 0.02 & 11.78 \\ 
  2 & 4.85 & 17.26 & 1.95 & 0.01 & 21.14 \\ 
  3 & 0.10 & 1.45 & 0.07 & 0.01 & 0.25 \\ 
  4 & 0.43 & 1.21 & 0.21 & 0.02 & 2.00 \\ 
  5 & 0.06 & 0.12 & 0.05 & 0.02 & 0.15 \\ 
  6 & 0.10 & 1.31 & 0.04 & 0.03 & 0.25 \\ 
  7 & 0.09 & 1.28 & 0.05 & 0.03 & 0.22 \\ 
\hline
\end{tabular}
\label{tab:duration}
\end{center}
\end{table}

%%=====================================================%%
% Fig: Duration
%%=====================================================%%

%%=====================================================%%
%%=====================================================%%
\section{Discussion}\label{sec:discussion}
% \textcolor{red}{Needs link back to Introduction/Dashboard Sections...}\\
Examining and viewing MES data in different ways leads to insights relating to the performance of the production line as a whole. As with all big data projects, a major challenge of this project was gaining an understanding of the characteristics of the data. It is essential to work closely with subject matter experts who can contextualize the data, and explain what the variables and the values they contain mean. Without this knowledge and context, the generation of insights (i.e.,~in a wholly automatic fashion) is much less feasible.

The main insights in our application have been in relation to workload, scraps, process step duration, and idle time --- quantities that would typically reside within MESs and would be of interest to business managers. In particular, the process steps with the most scraps and delays can be identified. The workload over time can be inspected across steps, with the aim of identifying shift times that are more variable. Action can be taken with the ultimate aim of optimizing production. In our example MES data, all process steps appear to behave similarly on average throughout time in relation to the number of starts, but there are differences in terms of the steps most likely to experience scraps or delays. The idle time values where units are not in production are highly skewed with large outlying values. Targeting and decreasing the time a unit is idle could improve throughput.

The effectiveness of the methods presented here are for indicative purposes only and would need to be further evaluated for daily in-field use. An alternative approach to the moving average window, such as cumulative sum charts and their confidence intervals, would provide additional insights into the detection of deviations from the expected process behavior. Moreover, root cause investigations would help to identify why one shift outperforms another, or, for example, why there are more units produced at a certain time on one day compared to another day. An additional layer would be to model the process steps with a view to predicting what will happen in the near future (e.g.,~30 minutes or one hour), or to predict the main drivers of events such as scraps and delays. The template day developed in this analysis can be improved by using ARIMA time series modeling \cite{arima}, for example, there may be monthly effects that are missed by simply averaging over all Wednesdays in the year.

The concept of the dynamic dashboard using today's data allows visualizations to be displayed live on screens in the production lines. Performance targets, such as reducing idle time by 10\%, can be set for the operators to achieve. The visualizations themselves would have to be developed in collaboration with the end users (e.g.,~process engineers) to ensure that the information is optimally conveyed, and deploying the tools to provide live feedback may provide some practical IT challenges. Logistics relating to where the visualizations are displayed and automatically running the code to produce the visualizations with the most recent data need to be considered.

The visualizations of the results illustrate the importance of carrying out an exploratory analysis. The visualizations and interactive dashboards provide evidence of process variability between shifts and days of the week. Exploiting insights from data is more reliable than anecdotal evidence. The potential value of the MES data is showcased, highlighting the many advantages of developing and utilizing process visualizations.

%%=====================================================%%
% Table: Color Rules
%%=====================================================%%
\begin{table}[t!]
\caption{Color rules based on percentiles}
\begin{center}
\begin{tabular}{|c|c|c|}
\hline
\textbf{Color}&\textbf{\# Start, \# Complete}&\textbf{\# Scrap, Idle Time, Duration}\\
\hline
Green & $\text{today} > {40}^\text{th}$ & $\text{today} < {60}^\text{th}$ \\ 
  Orange & ${20}^\text{th} < \text{today} < {40}^\text{th}$ & ${60}^\text{th} < \text{today} < {80}^\text{th}$ \\
  Red & $\text{today} < {20}^\text{th}$ & $\text{today} > {80}^\text{th}$\\
\hline
\end{tabular}
\label{tab:color_rules}
\end{center}
\end{table}

%%=====================================================%%
%%=====================================================%%
% \section*{References}

\end{document}